\documentclass[12pt]{article}
\usepackage{graphicx}
\usepackage{longtable}

\begin{document}

% 4 rows Astronomy Letters style:
\makeatletter
\renewcommand*{\@cite}[2]{{#2}}
\renewcommand*{\@biblabel}[1]{#1.\hfill}
\makeatother

\title{Comparison of the Pulkovo Compilation of Radial Velocities with the RAVE DR1 Catalogue}
\author{G.~A.~Gontcharov\thanks{E-mail: georgegontcharov@yahoo.com}}

\maketitle

Pulkovo Astronomical Observatory, Russian Academy of Sciences, Pulkovskoe sh. 65, St. Petersburg, 196140 Russia


The Data Release 1 of the Radial-Velocity Experiment (RAVE DR1, 24 748 stars)
is compared with the May 15, 2006 version of the Pulkovo Compilation of Radial Velocities (PCRV, 35 495 stars).
RAVE DR1 includes mostly $9^m-13^m$ stars, while the PCRV contains brighter stars. Analysis of the ``RAVE
minus PCRV'' radial-velocity differences for 14 common stars has revealed no systematic dependences
on any factors, except the effect due to the RAVE radial-velocity zero-point offset known from the RAVE
observations. This effect shows up for ten of these stars observed on a single night as a sine wave with an
amplitude of 1.5 km s$^{-1}$ in the dependence of the radial-velocity difference on the ordinal number of the
optical fiber used and, accordingly, on the star position angle in the field of view of the RAVE instrument.
The detection of this dependence confirms a high radial-velocity accuracy in both catalogs: on average,
better than 1 km s$^{-1}$ for stars brighter than $10^m$ (for the RAVE, after applying a correction for the zeropoint
offset). The RAVE zero-point offset can be corrected for with an accuracy better than 1 km s$^{-1}$ by
observing several PCRV stars in each RAVE frame and by analyzing the ``RAVE minus PCRV'' radialvelocity
differences.

\newpage
\section*{INTRODUCTION}

Two major radial-velocity catalogs were published
in 2006. Their comparison can improve the accuracy
of further radial-velocity observations.

The Pulkovo Compilation of Radial Velocities
(PCRV) whose May 15, 2006 release was presented
by Gontcharov (2006) contains 35 495 stars
brighter than $V=13^m$ from the Hipparcos catalogue
(ESA 1997). The radial velocities of these stars were
collected in the PCRV from more than 200 publications,
including the major ones: the WEB catalog
(Duflot et al. 1995), the catalogue by Barbier-Brossat
and Fignon (2000), the Geneva–Copenhagen Survey
of the solar neighborhood (below referred to as the
GCS) (Nordstr\"m et al. 2004), and the kinematic
survey of K and M giants (Famaey et al. 2005).
The observations of standard IAU stars or stars
from the compiled work list of secondary standards
are used to allow for the zero-point discrepancies
between the publications and the list of standards
and for significant systematic dependences of the
``publication minus list of standards'' radial-velocity
differences on the radial velocity itself, $(B-V)$ color
index, and equatorial coordinates. Particular attention
is given to the analysis of the radial velocities from the
four mentioned major publications. The discrepancy
between their zero points that was revealed when
compared with the list of standards does not exceed
0.5 km s$^{-1}$ and significant systematic dependences,
which do not exceed 0.4 km s$^{-1}$ in absolute value,
were found only in the GCS. Thus, combining these
data into a unified compilation (after applying the
appropriate corrections) is quite justifiable.

We used the standard deviation of the ``publication
minus list of standards'' radial-velocity differences
after applying a correction for the systematic
dependences and zero-point discrepancies to assign
weights and to calculate the weighted mean radial
velocity of each star and its accuracy. The median
accuracy of the weighted mean radial velocity in the
PCRV is 0.7 km s$^{-1}$. The radial velocities of 21 015
stars (59\%) have an accuracy better than 1 km s$^{-1}$.
However, unfortunately, the PCRV (like all of the
compilations) contain mostly stars whose radial velocities
were measured in only one observing program
and are given in only one publication: 24 437 stars
(69\%). Therefore, comparing the results of independent
radial-velocity measurements with different instruments
is of great importance.

One of the major projects in the history of astronomy
to measure the radial velocities of stars, the
Radial Velocity Experiment (RAVE) (Steinmetz et al.
2006), was started in 2003. By 2011, the spectra of
up to one million stars are to be taken and their radial
velocities and stellar atmosphere parameters are to be
determined. The Data Release 1 (DR1) of the catalog
of radial velocities was published in 2006. It includes
24 748 southern-sky stars mostly with magnitudes
$9^m-13^m$, about half of which enter the Tycho-2 catalogue
(H\o g et al. 2000).

The observations are performed with the spectrograph
mounted on the 1.2-m UK Schmidt Telescope
at the Anglo-Australian Observatory. The CCD array
simultaneously records the spectra of many objects
in a 6$^{\circ}$ field of view. The spectral lines in a narrow
infrared spectral range (841--880 nm) around the Ca
triplet are measured. The spectra have a mean resolution
of $R=7500$ and the pixel size corresponds,
on average, to a velocity of 13 km s$^{-1}$. Therefore,
the radial velocity can be measured theoretically with
an accuracy of at least 1.3 km s$^{-1}$ (1/10 pixel) at
a high signal-to-noise ratio. It is this scatter that is
achieved in the best cases during repeated observations
of stars brighter than $10^m$ at a signal-to-noise
ratio larger than 75. The accuracy can theoretically
be about 2 km s$^{-1}$ with a mean signal-to-noise ratio
of about 30 for the observed stars and will deteriorate
greatly at a lower signal-to-noise ratio for $12^m-13^m$
stars. However, the most important source of errors is
the radial-velocity zero-point offset reaching several
km s$^{-1}$ in one hour of observations that was revealed
by the RAVE team. If this offset is disregarded, then
it enters completely into the error of the measured
radial velocity and manifests itself when the radial
velocities of common PCRV and RAVE stars are
compared. Therefore, below we consider the parts of
the instrument and the peculiarities of the method for
determining the radial velocities in RAVE that can
have a bearing on the zero-point offset.

\section*{THE RAVE RADIAL-VELOCITY ZERO-POINT OFFSET}

In the RAVE instrument, light falls on a plate at
the focus of the telescope and is transmitted to the
stationary CCD spectrograph by 150 movable optical
fibers whose ends are located near this plate. In the
parked state, the fiber ends are distributed almost
uniformly in the form of a ring along the edge of the focal
plate. During measurements, an automatic device
alternately displaces the fiber ends from the edge of
the field of view to the star images. The displacement
line should not deviate from the radial direction by
more than 14$^{\circ}$. Therefore, each fiber is pointed at
objects predominantly in the sector of the field of view
nearest to it. It takes about one hour to record the
spectra of several tens of stars within the same 6$^{\circ}$ field
of view using most of the fibers together with the flatfield
spectra and the comparison spectra from a neon
arc before and after the star spectra were recorded.
The fibers are used in turn in order of increasing
number. Let us call all of the corresponding information
recorded by the CCD array for a single 6$^{\circ}$ field
a frame. The radial velocities are determined by the
standard cross-correlation method (Tonry and Davis
1979). Since stars of all spectral types and luminosity
classes are observed in the RAVE project, the computer
spectral ``mask'' used in the cross-correlation
method is formed based on the major libraries of theoretical
spectra by Zwitter et al. (2004) and Munari
et al. (2005). A total of 22 992 theoretical spectra were
used to compile the RAVE DR1 catalogue. As was
shown in the description of the RAVE DR1 catalogue,
the multistep procedure for choosing the best
theoretical spectrum for a specific star ensures a high
accuracy of radial-velocity measurements.

The RAVE team found that the emission lines in
the comparison spectrum from a neon arc after the
spectra of stars in the same frame were recorded are
shifted along the axes of the spectra relative to their
positions before the stellar spectra were recorded.
Special studies showed a correlation between this
shift and the change in the temperature of the instrument,
but the ultimate cause of this effect is unclear.
This shift results in a systematic radial-velocity zeropoint
offset reaching 5 km s$^{-1}$ within the same frame.
Since the comparison spectrum cannot be recorded
more often than two times for the same frame, the
RAVE team used the following procedure. The nightsky
emission lines recorded in the frame as a background
of stellar spectra separately for each optical
fiber are used to estimate the zero-point offset when
a single frame is recorded. The dependence of the
emission-line radial velocity on the ordinal number
of the optical fiber obtained in this way is then considered
as the zero-point offset, because the fibers
within the same frame are used in order of increasing
number. This dependence is fitted by a low-degree
polynomial and is subtracted from the stellar radial
velocities. Each frame has its own polynomial. Figure
1 shows examples of the zero-point corrections
as a function of the ordinal number of the optical fiber
for three frames on different nights. Unfortunately, the
signal-to-noise ratio is low for the night-sky emission
lines. Therefore, this method of correcting for
the zero-point offset has an accuracy no better than
1 km s$^{-1}$.

\section*{COMPARISON OF THE PCRV AND RAVE RADIAL VELOCITIES}

Figure 2 shows the distribution of PCRV and
RAVE DR1 stars in J magnitude from the 2MASS
catalog (we chose the J magnitude, because it is
known directly from observations almost for all of
the stars under consideration). The catalogues span
different magnitude ranges and have only 14 common
stars. This is clearly insufficient to fully reveal the systematic
errors in the RAVE results, but is quite sufficient
to detect the main systematic error in RAVE,
the zero-point offset, and, accordingly, to reach the
conclusion that regular observations of stars brighter
than 10m with well-known radial velocities are needed
in the RAVE project.

The ``RAVE minus PCRV'' radial-velocity differences
for 14 common stars show no significant systematic
dependences on magnitude, radial velocity,
$(B-V)$ color index, signal-to-noise ratio, height and
width of the peak of the correlation function, focal
plate number, and other factors, except the ordinal
number of the optical fiber.

Three of the 14 common stars are scattered
in different frames. The ``RAVE minus PCRV''
radial-velocity differences for them are 0.3, 1.1, and
3.9 km s$^{-1}$. The first two values agree with the estimated
accuracy in the catalogues. The last value was
obtained for the faintest of the 14 stars, HIP 67181,
with $V=10^m$, a high radial velocity (408 km s$^{-1}$,
RAVE), a peculiar spectrum, and presumed binarity
(Hipparcos DMSA stochastic solution). The remaining
11 stars were observed in two successive frames
on August 6, 2003. Table 1 gives parameters of these
stars: Hipparcos numbers, approximate coordinates
$\alpha$ and $\delta$ (J2000) in degrees with fractions, V magnitudes,
and $(B-V)$ color indices from Hipparcos.
Table 2 gives data referring to the radial velocities
of these stars: Hipparcos numbers, radial-velocity
sources (only one for all stars, except HIP 78432,
for which two sources were used in the PCRV),
radial velocities from the source (for HIP 78432,
the mean of two sources), PCRV radial velocities,
RAVE DR1 radial velocities, frame designations,
signal-to-noise ratios, and the height of the peak of
the correlation function in the RAVE observations.
Since these frames refer to neighboring regions of the
sky and were recorded successively, without any gaps
in time, given the correlation between the zero-point
offset and the temperature revealed by the RAVE
team, a similar zero-point offset can be assumed in
these frames. No correction for the zero-point offset
was applied in the RAVE DR1 catalogue for these
frames. Therefore, the zero-point offset shows up
as the dependence of the ``RAVE minus PCRV''
radial-velocity differences on the ordinal number of
the optical fiber. These differences together with the
``RAVE minus GCS'' differences (in km s$^{-1}$) are
indicated in Fig. 3 as a function of the ordinal number
of the optical fiber by the diamonds and squares,
respectively. The vertical bars indicate the accuracy of
the PCRV radial velocities, which is almost equal to
that of the GCS radial velocities. It makes no sense to
give the accuracy from the RAVE, because the zeropoint
offset was disregarded when it was calculated.

The ``RAVE minus PCRV'' radial-velocity differences
for the 10 stars mentioned above (7 in one
frame and 3 in the other frame) fall nicely on the
$V_{RAVE}-V_{PCRV}-1.50\sin(\theta-45^{\circ})+0.58$ km s$^{-1}$
curve, where $\theta$ is the position angle of the optical fiber
in the field of view (in degrees). The star HIP 79 777
gave an outlier (encircled in Fig. 3), which may result
from the smallest height of the peak of the correlation
function (0.88) among the 14 stars. The sine wave
found resembles the polynomials for other RAVE
frames presented in Fig. 1 as examples, but is was
drawn much more accurately. Figure 4 shows the
correlation between the radial-velocity differences
and $\sin(\theta-45^{\circ})$ for 10 stars (without HIP 79 777):
the correlation coefficient is 0.94. The correlation
coefficient for the GCS radial velocities of eight stars
is 0.93 (HIP 79777 also gives an outlier in the ``RAVE
minus GCS'' difference). We see that the GCS radial
velocities were included in the PCRV with small
corrections. Thus, the authors of RAVE DR1 had
sufficient information (the radial velocities of nine
GCS stars) to detect the zero-point offset from the
``RAVE minus GCS'' differences.

If we take the PCRV zero point, then a tentative
(due to the small number of stars) conclusion about
the necessary mean RAVE radial-velocity zero-point
correction, about $-0.58$ km s$^{-1}$, can be drawn from
the mean ``RAVE minus PCRV'' radial-velocity difference.
This value agrees well with the mean zeropoint
correction of $-0.55$ km s$^{-1}$ determined by the
RAVE team from night-sky emission lines, which
is indicative of a high accuracy of allowance for the
zero-point offset on average. However, an accurate
allowance for the zero-point offset is needed for each
frame as well. For example, for the frames in question,
allowance for the sine wave found reduces the standard
deviation of the radial-velocity differences for
these 10 stars from 1.2 to 0.4 km s$^{-1}$. This is probably
the real mean accuracy of the radial velocities for stars
brighter than 10m in the PCRV and RAVE DR1,
which is quite achievable after applying an accurate
correction for the zero-point offset.

\begin{table}
\def\baselinestretch{1}\normalsize\small
\caption[]{Parameters of 11 common PCRV and RAVE DR1 stars.
}
\label{stars1}
\[
\begin{tabular}{ccccc}
\hline
\noalign{\smallskip}
 HIP & $\alpha$ & $\delta$ & V & $(B-V)$ \\
\hline
\noalign{\smallskip}
78432 & 240.18 & -47.89 & 8.7 & 1.39 \\
78791 & 241.27 & -48.83 & 8.3 & 0.59 \\
79777 & 244.22 & -49.86 & 7.6 & 0.76 \\
79912 & 244.67 & -51.21 & 8.7 & 0.52 \\
83415 & 255.75 & -43.12 & 9.5 & 0.85 \\
84152 & 258.07 & -43.49 & 8.5 & 0.73 \\
84419 & 258.88 & -42.78 & 8.5 & 0.60 \\
84911 & 260.30 & -41.56 & 8.7 & 0.74 \\
85211 & 261.18 & -45.01 & 6.7 & 0.39 \\
85222 & 261.21 & -41.50 & 7.8 & 0.52 \\
85468 & 261.98 & -42.41 & 8.4 & 0.62 \\
\hline
\end{tabular}
\]
\end{table}

\begin{table}
\def\baselinestretch{1}\normalsize\small
\caption[]{Radial velocities of 11 common PCRV and RAVE DR1 stars
(DCZ stands for Da Costa and Seitzer (1989).}
\label{stars2}
\[
\begin{tabular}{cccrrrcc}
\hline
\noalign{\smallskip}
 HIP & Source & $V_r$, km s$^{-1}$ & PCRV, km s$^{-1}$ & RAVE, km s$^{-1}$ & Frame & S/N & Peak \\
\hline
\noalign{\smallskip}
78432 & WEB/DCZ & -20.9 & -20.7 & -18.0 & 1607m49 &  61 & 0.97 \\
78791 & GCS &  -0.3 &  -0.1 &   1.7     & 1607m49 &  55 & 0.96 \\
79777 & GCS &  10.8 &  11.2 &  14.2     & 1607m49 & 117 & 0.88 \\
79912 & WEB &  -6.2 &  -6.2 &  -7.2     & 1607m49 &  52 & 0.95 \\
83415 & GCS &  -6.8 &  -6.3 &  -4.7     & 1716m42 &  57 & 0.94 \\
84152 & GCS & -49.7 & -49.3 & -49.3     & 1716m42 &  75 & 0.95 \\
84419 & GCS &  -2.4 &  -2.2 &  -0.3     & 1716m42 &  50 & 0.94 \\
84911 & GCS & -52.5 & -52.1 & -51.4     & 1716m42 &  72 & 0.95 \\
85211 & GCS &  -1.3 &  -1.7 &  -1.9     & 1716m42 & 147 & 0.95 \\
85222 & GCS &  24.8 &  24.8 &  24.5     & 1716m42 & 133 & 0.96 \\
85468 & GCS & -20.7 & -20.4 & -20.7     & 1716m42 &  59 & 0.94 \\
\hline
\end{tabular}
\]
\end{table}

The actually considered effect is a blue and red
shift of the spectral lines at the opposite edges of the
focal plate (in the frames under consideration, at the
``southeastern'' and ``northwestern'' edges). In this
case, the zero point may change not just because of
the temperature changes, but because these changes
compound a certain dependence of the zero point
on the position angle in the field of view. A similar
effect was found when analyzing the focal scale of
the photographic vertical circle (PVC) at the Pulkovo
Observatory of the Russian Academy of Sciences)
(Gontcharov 1996). Since the focal photographic
plate is inclined to the PVC optical axis, the scale
changes as $r(at+b\sin\theta)$, where $\theta$ is the position
angle in the field of view, t is the time, r is the distance
from the star to the intersection of the optical axis
with the focal plane, and a and b are temperaturedependent
coefficients. The RAVE zero-point offsets
shown in Fig. 1 can be represented by a similar formula.

Special studies would probably allow the radial velocity
zero-point offset in the RAVE observations
to be taken into account more accurately. It may be
useful to control the orientation of the focal plate
relative to the optical axis. Undoubtedly, observations
of radial-velocity standards and/or PCRV stars will
be useful. Given the PCRV star density (at least one
star with an accurate radial velocity per two square
degrees) and their uniform sky distribution, several
PCRV stars can be recorded in each frame of the
6$^{\circ}$ field by different optical fibers at different position
angles and the zero-point offset can be taken into
account for each frame with an accuracy no worse
than the mean accuracy of the PCRV radial velocities,
0.7 km s$^{-1}$, i.e., much more accurately than is
now done using night-sky emission lines, by analyzing
the ``observations minus PCRV'' radial-velocity
differences.

\section*{CONCLUSIONS}

When the two major radial-velocity catalogs are
compared, the main systematic error of the RAVE
observations, the zero-point offset, manifests itself in
the ``RAVE minus PCRV'' radial-velocity differences.
This suggests that the typical accuracy of the radial
velocities for stars brighter than $10^m$ in both catalogues
is appreciably better than 1.5 km s$^{-1}$, the
amplitude of the detected systematic effect, and, consequently,
it is no worse than the accuracy declared
by the authors of both catalogs. The accuracy of the
RAVE data can be improved further (probably up to a
level of 1 km s$^{-1}$ for several hundred thousand Tycho-
2 stars brighter than $11^m$) by a careful allowance
for the zero-point offset, for example, using regular
observations of standard stars or PCRV stars.

Compared to the PCRV, the mean RAVE radial velocity
zero-point correction ($-0.58$ km s$^{-1}$) agrees
well with the corresponding correction determined
by the RAVE team from night-sky emission lines:
$-0.55$ km s$^{-1}$. This is indicative of a high accuracy
of allowance for the zero-point offset in the RAVE
results \emph{on average}.

\section*{ACKNOWLEDGMENTS}

I used the RAVE Data Release 1 (http://www.ravesurvey.aip.de/rave/)
and resources of the Astronomical
Data Center in Strasbourg (France) (http://cdsweb.u-strasbg.fr/).
This work was supported by
the Russian Foundation for Basic Research (project
no. 05-02-17047).

\newpage

\begin{figure}
\includegraphics{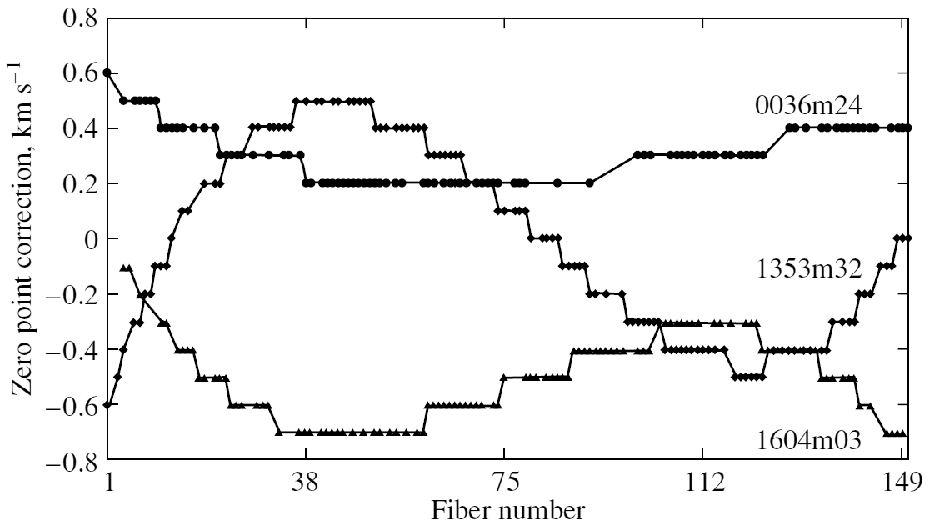}
\caption{Examples of the radial-velocity zero-point corrections (in km s$^{-1}$) determined from night-sky emission lines
and applied to the RAVE observations as a function of the ordinal number of the optical fiber in the field of view for
several frames on different nights: 0036m24, 1353m32, 1604m03.}
\label{zero}
\end{figure}

\begin{figure}
\includegraphics{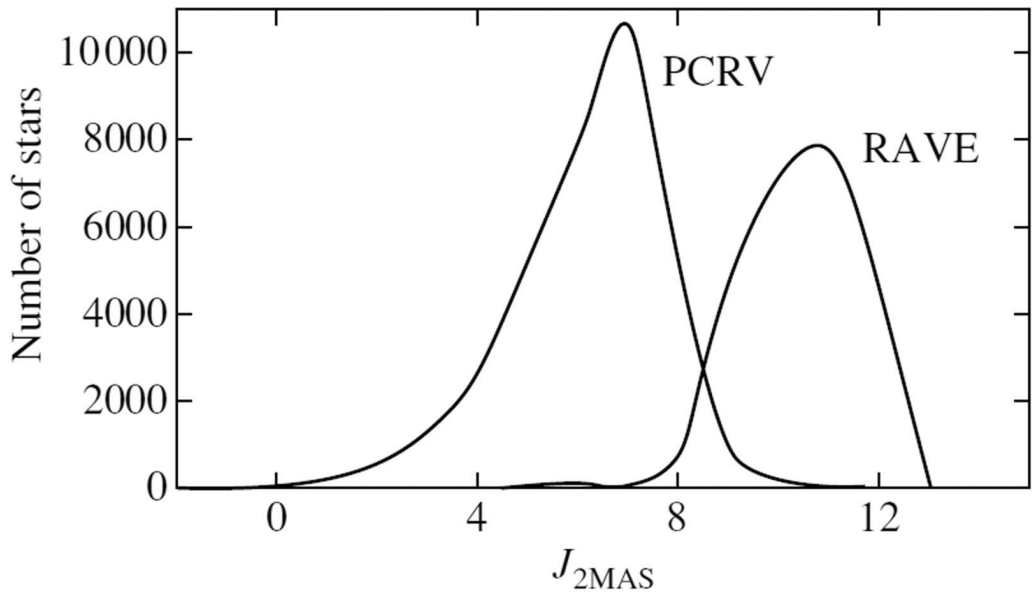}
\caption{Distribution of PCRV and RAVE DR1 stars in J magnitude from the 2MASS catalog.}
\label{dist}
\end{figure}

\newpage

\begin{figure}
\includegraphics{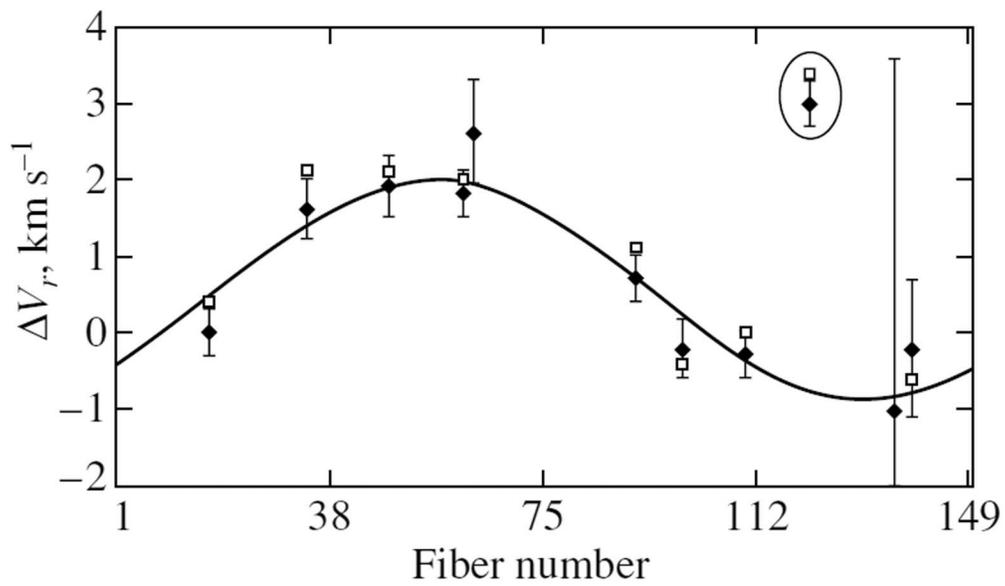}
\caption{``RAVE minus PCRV'' (diamonds) and ``RAVE minus GCS'' (squares) radial-velocity differences vs. ordinal
number of the optical fiber in the RAVE field of view for 11 stars observed on August 6, 2003. The vertical bars
indicate the accuracy from the PCRV. The outlier for the star HIP 79777 is encircled.}
\label{main}
\end{figure}

\begin{figure}
\includegraphics{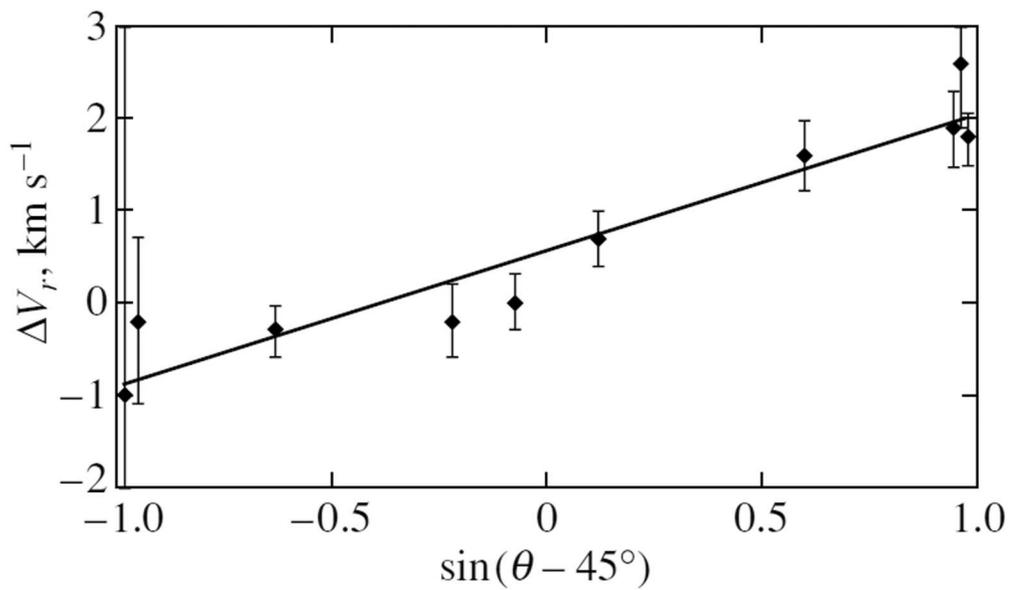}
\caption{``RAVE minus PCRV'' radial-velocity differences vs. $\sin(\theta-45^{\circ})$, where $\theta$ is the position
angle of the optical fiber in the field of view (in degrees) for 10 common
stars observed on August 6, 2003. The vertical bars indicate the accuracy from the PCRV.}
\label{corr}
\end{figure}

\end{document}